%% file: main.tex
\documentclass[reprint,amsmath,amssymb,aps]{revtex4-2}

\usepackage{dcolumn}
\usepackage{bm}
\usepackage{graphicx}
\usepackage[utf8]{inputenc}
\usepackage{csquotes}
\usepackage[english]{babel}
\usepackage[pdftex, pdftitle={Article}, pdfauthor={Author}]{hyperref} 
\usepackage[yyyymmdd]{datetime}

\usepackage{subcaption}
\usepackage{newunicodechar}
\DeclareFontEncoding{LS1}{}{}
\DeclareFontSubstitution{LS1}{stix}{m}{n}
\DeclareFontFamily{LS1}{stixscr}{\skewchar\font127 }
\DeclareFontShape{LS1}{stixscr}{m}{n} {<->s*[1.2] stix-mathscr}{}
\newunicodechar{◌}{{\usefont{LS1}{stixscr}{m}{n}\symbol{\string"E3}}}
\usepackage{wasysym}
\usepackage{tikz,scalerel}
\usetikzlibrary{shapes.geometric}

\newcommand\polygon[3][]{%
    \begin{scope}[#1]%
       \draw (-90+360/#2:#3)%
       foreach \i in {1,2,...,\the\numexpr#2-1}{%
            -- (-90+360/#2 +\i*360/#2:#3)}%
    -- cycle;%
    \end{scope}%
}%

\def\pentagondown{\scalerel*{\begin{tikzpicture}%
      \polygon{5}{1};%
\end{tikzpicture}}}

\def\pentagonup{\scalerel*{\begin{tikzpicture}%
    \node[regular polygon,draw] (p) at (0,0) {};%
\end{tikzpicture}}{()}}%

\begin{document}

\title{Complex motion of steerable vesicular robots filled with active colloidal rods}

\author{Sophie Y. Lee}
    \affiliation{Department of Materials Science and Engineering, University of Michigan, Ann Arbor, Michigan 48109, USA}

\author{Philipp W. A. Sch\"{o}nh\"{o}fer}
    \affiliation{Department of Chemical Engineering, University of Michigan, Ann Arbor, Michigan 48109, USA}

\author{Sharon C. Glotzer}
    \email[Correspondence email address: ]{sglotzer@umich.edu}
    \affiliation{Department of Materials Science and Engineering, University of Michigan, Ann Arbor, Michigan 48109, USA}
    \affiliation{Department of Chemical Engineering, University of Michigan, Ann Arbor, Michigan 48109, USA}
    \affiliation{Biointerfaces Institute, University of Michigan, Ann Arbor, Michigan 48109, USA}

    \date{\formatdate{20}{6}{2023}} 

\begin{abstract}
    While the collective motion of active particles has been studied extensively, effective strategies to navigate particle swarms without external guidance remain elusive. We introduce a method to control the trajectories of two-dimensional swarms of active rod-like particles by confining the particles to rigid bounding membranes (vesicles) with non-uniform curvature. We show that the propelling agents spontaneously form clusters at the membrane wall and collectively propel the vesicle, turning it into an active superstructure. To further guide the motion of the superstructure, we add discontinuous features to the rigid membrane boundary in the form of a kinked tip, which acts as a steering component to direct the motion of the vesicle. We report that the system's geometrical and material properties, such as the aspect ratio and P\'eclet number of the active rods as well as the kink angle and flexibility of the membrane, determine the stacking of active particles close to the kinked confinement and induce a diverse set of dynamical behaviors of the superstructure, including linear and circular motion both in the direction of, and opposite to, the kink.  From a systematic study of these various behaviors, we design vesicles with switchable and reversible locomotions by tuning the confinement parameters. The observed phenomena suggest a promising mechanism for particle transportation and could be used as a basic element to navigate active matter through complex and tortuous environments.
\end{abstract}

\maketitle

\section{Introduction}
The ongoing and intense interest in active matter inarguably involves its promise as a precursor to microscopic colloidal robots. If the emergent, collective motion of self-propelled particles central to active matter could be directed and harnessed for work, colloidal machines that locomote, capture and retrieve tiny objects, and other robotic functions could be achieved for applications ranging from \textit{in vivo} health diagnostics and drug delivery to stealth. However, despite the rapid, ongoing development of new synthesis techniques \cite{wang_practical_2020} and computational models \cite{shaebani_computational_2020,zottl_modeling_2023} of  particles that convert external energy into an internal driving force, their controlled and efficient collective transport to mimic robotic entities still faces major technical challenges.  Such challenges include the robustness of active swarm navigation against thermal noise and other perturbations, enabling switchable and reversible mechanisms to direct transport, implementing programmable sensing, and triggering responses to external stimuli and complex environments  \cite{gompper_delivering_2022,gompper_2020_2020}. A variety of external control strategies have been applied to realize sensing and directed navigation \cite{nsamela_colloidal_2022}, such as the use of external fields \cite{arlt_painting_2018, merlitz_pseudo-chemotaxis_2020, fernandez-rodriguez_feedback-controlled_2020, singh_interface-mediated_2020-1} or an external feedback loop \cite{lavergne_group_2019,bregulla_stochastic_2014,yang_cargo_2020, muinos-landin_reinforcement_2021, yang_autonomous_2022}. In contrast to external control strategies, autonomous control strategies could be advantageous for many applications, but these strategies have received considerably less attention in the literature.\\

One such autonomous approach might be to design active particle superstructures that permit autonomous navigation as a result of their intrinsic and self-emergent behavior. Here we can draw inspiration from macroscopic robotic swarms. Particle robots with pre-programmed interactions can perform robotic tasks such as targeted morphological changes \cite{rubenstein_programmable_2014} or synchronized motion \cite{li_particle_2019}, but such high level of internal logic is not necessary. For example, stochastic robotic collectives are capable of global locomotion and are able to achieve complex goals by transforming task-incapable single components into task-capable robotic swarms \cite{savoie_robot_2019}. A promising example is that of macroscopic, forward-propelling, rod-like robotic units that self-organize when trapped inside a deformable confining boundary by accumulating at the boundary \cite{deblais_boundaries_2018, uplap_design_2023}, turning the system into a moving superstructure \cite{boudet_collections_2021, gao_self-driven_2017}. In an effort to engineer cellular migration, similar concepts have been reported in studies of shape modulation and transport dynamics of vesicular membranes driven by encapsulated active components \cite{le_nagard_encapsulated_2022, ramos_bacteria_2020, paoluzzi_shape_2016, abaurrea-velasco_vesicles_2019, li_shape_2019, vutukuri_active_2020,iyer_non-equilibrium_2022, kokot_spontaneous_2022-1, gao_self-driven_2017, sanchez_spontaneous_2012-1, iyer_dynamic_2023-2}. For instance, motile bacteria have been used to drive droplets inside an emulsion \cite{ramos_bacteria_2020} or to generate cell extrusion in lipid vesicles before they propel the vesicle forward \cite{le_nagard_encapsulated_2022}.\\

Inspired by these autonomous superstructure strategies, in this paper we expand on the idea of encapsulation of active particles as a device to control their collective motion by adding an asymmetric bias to the superstructure model. Bias is a known requirement for achieving asymmetric behavior in active matter systems \cite{di_leonardo_bacterial_2010, gentile_chemically_2020, gao_self-driven_2017, sanchez_spontaneous_2012-1}, and here we achieve asymmetry via the geometry of the confining boundary. Using molecular dynamics simulation (see methods section), we confine self-propelling rod-shaped particles inside rigid vesicles whose discontinuous curvature creates kinks in the vesicle boundary. We first show that the active rods tend to form clusters at the kink, similar to clusters of active particles observed in channels \cite{wensink_aggregation_2008,caprini_activity-controlled_2020,al_alam_active_2022}, under polygonal confinement \cite{frangipane_invariance_2019} or near a chevron-shaped trap \cite{kaiser_how_2012, reichhardt_ratchet_2017}. The resulting bias in the location of the emergent cluster leads to a tunable and directed propulsion of the vesicular superstructure. Based on the properties of the vesicle (rigidity and kink angle) and the active rods (P\'eclet number, aspect ratio, particle density inside the vesicle) we observe linear motion and circular motion, both in the direction of the kink (kink-forward) and opposite to the direction of the kink (kink-backwards). We relate these four different types of motion to the alignment and stacking of the active rods at the kink and showcase how this mechanism can be used to create vesicles with switchable and reversible dynamics.

\section{Method}

\begin{figure}[ht]
  \centering
  \includegraphics[width=0.8\linewidth]{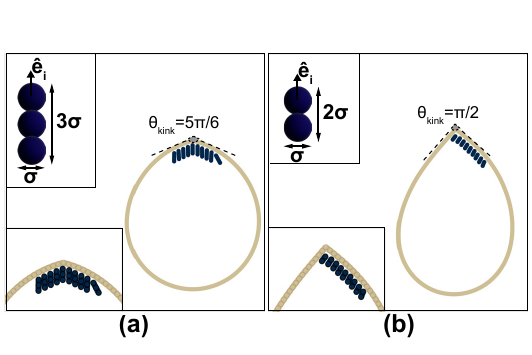}
  \caption{\label{fig:methods}\textbf{Active rods confined in a vesicle.} Sketch of a vesicle containing rigid active rods (dark blue) with (a) aspect ratio $\alpha=3$ aligning parallel to the long axis of the vesicle (beige) at a wide kink($\theta_{kink}=5\pi/6$), and (b) $\alpha=2$ aligning at an angle with one side of the vesicle wall (beige) at a narrow kink ($\theta_{kink}=\pi/2$). The upper inset shows the (a) trimer ($\alpha=3$) rod, and (b) dimer ($\alpha=2$) rod. The lower inset is a close-up view of the kinked part of the vesicle and the stacking of active rods where (a) has \textit{parallel} and (b) has \textit{angled} stacking.}
  \end{figure}
  
  We performed Brownian molecular dynamics simulations of self-propelled rod-shaped particles confined inside a vesicle in a square box with length $L_b = 125\sigma$. We modeled the vesicles as chains of $N_v$=100 bonded disks with diameter $\sigma=1$. The active rod-shaped particles are modeled as a rigid body of $m={2,3}$ disks with diameter $\sigma$ that are linked linearly end-to-end (see Fig \ref{fig:methods}) \cite{glaser_pressure_2020, nguyen_rigid_2011}. The number of active rods per vesicle perimeter varies between $\rho_{N_a}=\frac{N_a}{N_v\sigma}\in\{0.05,0.1,\dots,0.3\}$. We chose this definition of density to compare 2- and 3-rod systems that apply the same amount of force to the vesicle wall. The equations of motion of each vesicle disk and self-propelling rod is given by Brownian dynamics:
  \begin{align}
  \gamma \dot{\mathbf{r}_i} &= \sum_{j}\mathbf{F}_{ij}^{WCA} + F_{i}^{A}\hat{\mathbf{e}}_i + \mathbf{F}_i^{Ves} + \sqrt{2\gamma k_BT}\eta_i(t)\\
  \gamma_r \dot{\omega_i} &= \sum_{j}\mathbf{T}_{ij}^{WCA} +  \sqrt{2\gamma_r k_BT}\zeta_i(t)
  \end{align}
  with translational drag coefficient $\gamma$ = 1, and rotational drag coefficient $\gamma_r = \frac{\sigma^3\gamma}{3}$ per the Stokes-Einstein relationship. The interparticle forces $\mathbf{F}_{ij}^{WCA} $ on each disk and torques $\mathbf{T}_{ij}^{WCA}$ on each rod are derived from the purely repulsive Weeks-Chandler-Anderson potential
  \begin{align}
  U^{WCA}(r_{ij}) = 
  \begin{cases} 
    4\epsilon \left[ 
    \left( \frac{\sigma}{r_{ij}} \right)^{12} 
    - \left( \frac{\sigma}{r_{ij}} \right)^{6} 
    \right]+\epsilon & r_{ij} < r_c = \sqrt[6]{2}\sigma \\
    0 & r_{ij} > r_c,
  \end{cases}
  \end{align}
  where $r_{ij}$ is the distance between disk $i$ and $j$ and $r_c$ is the cut-off distance. The functions $\eta_i(t)$ and $\zeta_i(t)$ are normalized Gaussian white noise processes with zero mean $\langle \eta_i(t) \rangle=0$, $\langle \zeta_i(t) \rangle=0$ and unit variance $\langle \eta_i(t) \eta_j(t') \rangle = \delta_{ij}\delta(t-t')$, $\langle \zeta_i(t) \zeta_j(t') \rangle = \delta_{ij}\delta(t-t')$. The active force term $F_{i}^{A}\hat{\mathbf{e}}_i$ is only applied to the rod center and points along the symmetry axis of the rod $\hat{\mathbf{e}}_i=(\cos\omega_i,\sin\omega_i)$. The active force magnitude $F_i^{A}=\text{Pe}\cdot k_BT$ is controlled by the P\'eclet number $\text{Pe}\in\{25,50,75\}$ for the active rods and $F_i^A=0$ for the vesicle particles. The final term $\mathbf{F}_i^{Ves}$ combines the vesicle bond forces and, hence, is derived from the bond potentials 
  \begin{align}
  U_i^{Ves} &= \sum_{j} U_{FENE}(r_{ij}) +\sum_{k,l} U_{angle}(\theta_{ikl}),\\
  U_{FENE}(r_{ij}) &= -\frac{1}{2} \kappa_{b} r_0^2 \ln \left[ 1 - \left(\frac{r_{ij}}{r_0}\right)^2 \right],\\
  U_{angle}(\theta_{ikl}) &= -\frac{1}{2} \kappa_{a} (\theta_{ikl} - \theta_0 )^2.
  \end{align}
  Each set of neighboring disks $(i,j)$ comprising the vesicle boundary (membrane) is bonded by the finitely extensible nonlinear elastic (FENE) potential $U_{FENE}(r_{ij})$ \cite{warner_kinetic_1972} with bonding strength $\kappa_b=100k_BT\sigma^{-2}$ and equilibrium distance $r_0=2\sigma$. To apply a uniform bending rigidity to the vesicle membrane, we added a harmonic angle potential with bending coefficient $\kappa_{a}=1000k_BT$ and rest angle $\theta_0=\pi$ to every triplet $(ikl)$ of neighboring disks. We also ran several simulations with smaller bending coefficients  $\kappa_{a}\in\{50k_BT,125k_BT,250k_BT,500k_BT\}$ (Fig. S1), however, in the following all results are presented for rigid vesicle membranes with $\kappa_{a}=1000k_BT$ unless explicitly stated otherwise. Finally, we introduce a discontinuous element, which we refer to as a "kink", by setting the rest angle $\theta_0$ at one of the triplets to $\theta_0=\theta_{kink}\in\{\frac{\pi}{3},\frac{5\pi}{12},\frac{\pi}{2},\frac{7\pi}{12},\frac{2\pi}{3},\frac{3\pi}{4},\frac{5\pi}{6}\}$ (see Fig. \ref{fig:methods}).\\
  
  We ran each simulation \cite{towns_xsede_2014} at a temperature $k_BT=0.01$ for a time period $t=2\times10^4\tau$ with time step $\Delta t=1\times10^{-4}\tau$ and unit time $\tau = \sqrt{m\sigma^2/k_B T}$. To obtain proper statistics each point in the parameter space is sampled over 200 independent replica simulations. We used the open source molecular dynamics software HOOMD-blue \cite{anderson_hoomd-blue_2020} [v3.0.0] to perform our simulations,the  freud data analysis package\cite{ramasubramani_freud_2020} for cluster analysis and the signac software package \cite{adorf_simple_2018-1, dice_signac_2021-1} for data management.

  \section{Results}
  To illustrate the range of dynamical behaviors observed for the vesicle superstructure, we present their observed steady-state behavior as a function of kink angle $\theta_\text{kink}$ and rod particle density $\rho_{N_a}$ for both dimer and trimer rod systems in Fig. \ref{fig:results1} and SI Fig. S4. In all of our simulation we observe that the microscopic driving forces of the active agents are transferred to the vesicle in a coordinated manner, propelling the entire vesicle superstructure. Flexible or uniformly curved vesicles move in random and unpredictable directions (see Fig. \ref{fig:results1}) as seen in earlier studies \cite{uplap_design_2023}. However, at high bending rigidity $\kappa_a=1000k_BT$ and for non-uniform membrane curvature, the activity of the rods is transferred to the vesicle in a coordinated directed motion based on the position of the kink in the vesicle membrane. Overall, we obtain four different patterns of motion for kinked vesicles with different collective propelling mechanisms of the rod-shaped particles: a motion in the direction of the kink (\textit{linear forward}), a motion in the opposite direction of the kink (\textit{linear backward}), a kink \textit{forward circular} motion, in which the trajectories of the vesicles make loops, and a kink \textit{backward circular} motion (see Fig. \ref{fig:results1}). We describe each of the behaviors below.
  
  \subsection{Linear motion}
  
\begin{figure*}
  \centering
  \includegraphics[width=0.9\linewidth]{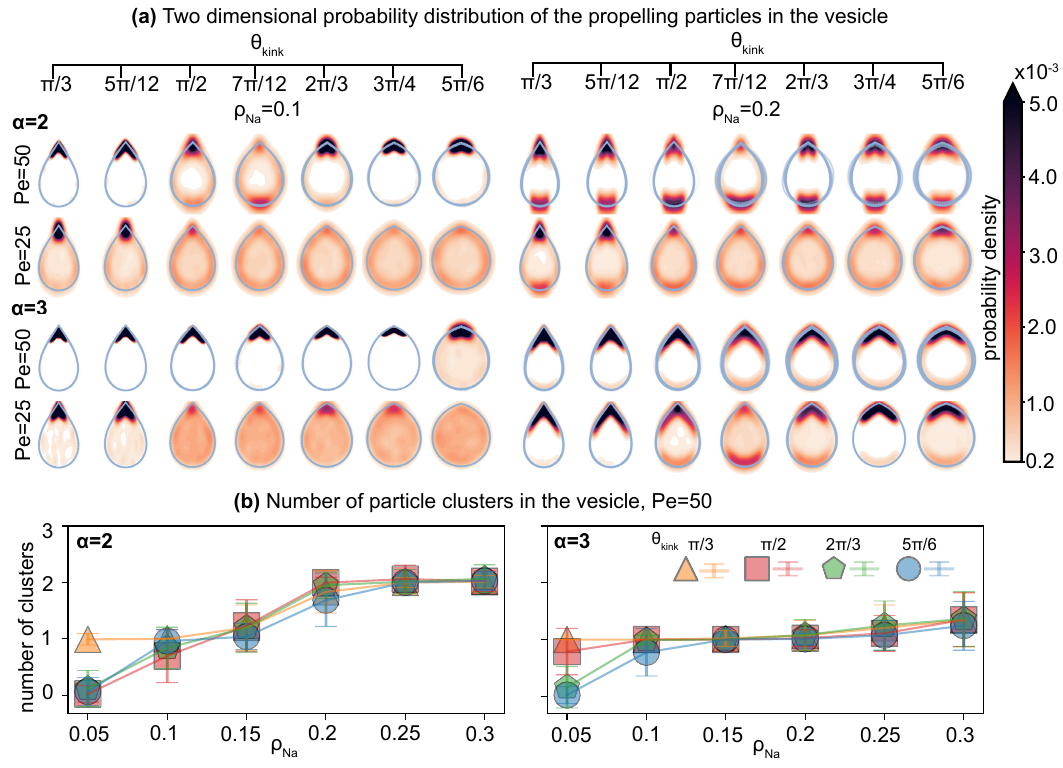}
  \caption{\label{fig:results2} \textbf{Spatial distribution of the self-propelled rods confined in an anisotropic vesicle.} \textbf{(a)} Two-dimensional probability distributions of the rods in a vesicle. The colors indicate the probability density per bin. \textbf{(b)} The average number of particle clusters in the vesicle as a function of increasing number density $\rho_{N_a}$ at different $\theta_{kink}$ for each aspect ratio $\alpha=2$ and $\alpha=3$. Each data point is averaged over 200 independent replicas. Every vesicle in the per-vesicle map \textbf{(a)} and in \textbf{(b)} has bending rigidity coefficient $\kappa_{a}=1000k_BT$. For other $\kappa_{a}$, see SI.}
  \end{figure*}

\begin{figure*}
  \centering
  \includegraphics[width=0.8\linewidth]{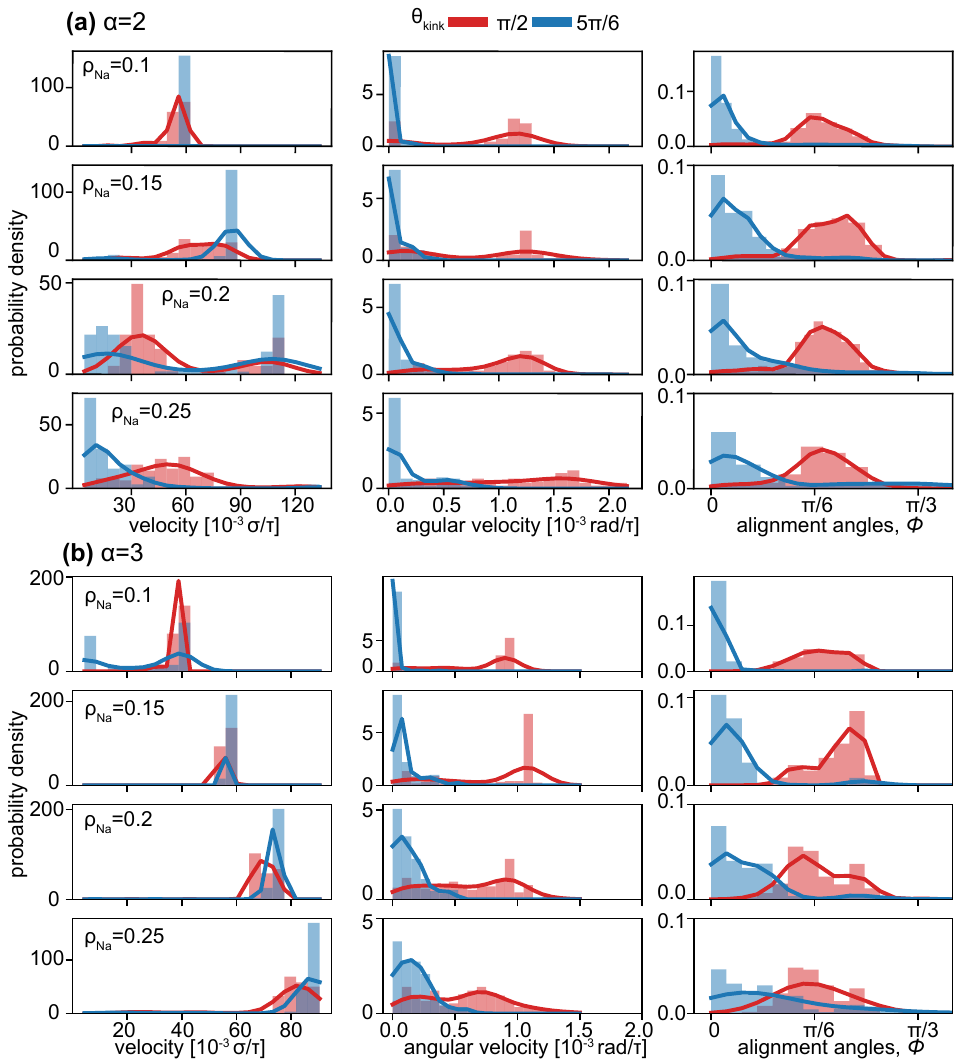}
  \caption{\label{fig:results3} \textbf{Histogram of the motion descriptors of a vesicle.} The histograms describe the distribution of two motion metrics (centroid velocity and angular velocity) and alignment angles of clustered particles as a function of $\theta_{kink}$ and $\rho_{N_a}$. The set of histograms are plotted for \textbf{(a)} $\alpha=2$, P\'{e}=75 and \textbf{(b)} $\alpha=3$, P\'{e}=50, respectively with $\theta_{kink}=\pi/2, 5\pi/6$. The first column indicates the velocity of the vesicle centroid. The second column indicates the angular velocity of the vesicle. The third column is the Kernel Density Estimation (KDE)\cite{noauthor_kernel_1992} approximation representing the distribution of alignment angles of clustered particles between the vesicle symmetry axis and the long axis of each active rod. Each row indicates the number density $\rho_{N_a}$ of active rods in the vesicle. Solid lines at the top of the histogram are KDE approximations of the corresponding $\theta_{kink}$ in matching colors. Each instance in the histogram corresponds to a single simulation and the metrics are computed and collected from the latter half of the simulation trajectory.}
  \end{figure*}

  Vesicles driven by rigid dimers (Fig.~\ref{fig:results1}a) or trimers (Fig.~\ref{fig:results1}b) predominantly exhibit linear motion at high $\theta_{kink}$. We identify that the linear trajectories of vesicles are correlated to a \textit{parallel} stacking (with respect to the long axis of the vesicle) of the active particles inside the vesicle near the kinked tip (see sketch in Fig.~\ref{fig:methods}a). In this regime, the rod-shaped particles first aggregate at the vesicle boundary individually and move along the interface as reported in other studies of active matter in hard \cite{deblais_boundaries_2018, giomi_swarming_2013} and flexible confinement \cite{nikola_active_2016, boudet_collections_2021}. Due to the discontinuity in curvature, which is known to slow down and capture active particles \cite{kaiser_how_2012}, the rods accumulate into dense stackings at the kink, align with the kink, and apply collective local pressure to the vesicle boundary in the direction of the rod alignment director. At low particle densities $\rho_{N_a} < 0.15$ the density distribution inside the vesicles indicates that the clusters are stable at the kink in steady state (see Fig.~\ref{fig:results2}a). Furthermore, the director of the aligned clusters of both dimers and trimers is parallel to the symmetry axis of the vesicle (see alignment angle distributions in Fig.~\ref{fig:results3}), with particles distributed roughly evenly on both sides of the kink (see Fig.~\ref{fig:methods}a). As a consequence, the vesicle superstructures are pushed forward with a small variance in velocity and hardly any rotational component to its locomotion (see velocity and angular velocity distribution in Fig.~\ref{fig:results3}).\\
  
  We find that the propulsion mechanism breaks down if the self-propelling force is too weak ($Pe < 50$) (see Fig.~\ref{fig:results2}a), the vesicle membrane is too flexible ($\kappa_a < 250k_BT$), or the perimeter of the vesicle exceeds 200$\sigma$. In the former case, the rods are less likely to cluster, resulting in a more uniform particle distribution inside the vesicle (see Fig.~\ref{fig:results2}a) and a lower yield of forward-moving vesicles. Additionally, clusters that do emerge at the kink are prone to dissolving again due to the decrease in active particle pressure. For flexible vesicles, we must consider shape fluctuations of the membrane, which are otherwise negligible for the highest bending coefficient $\kappa_{a}=1000k_BT$ (see Fig. S1). With decreasing $\kappa_{a}$, the pressure applied to the vesicle walls by the active rods produces local deformations and high curvature regions at the vesicle boundary. This creates additional nucleation pockets for clustering that can capture more rods, diminishing or eliminating the benefit of the kink. We also investigated different vesicle perimeter lengths between $110 \sigma$ and $500\sigma$ with monomer diameter $\sigma$. We observed that the resulting circular and linear motion of the vesicular superstructure guided by $\theta_{kink}$ gradually start to fail when the perimeter exceeds 200$\sigma$. In these larger vesicles, the effect that active particles spontaneously form crystallites away from the kink and produce additional pockets of local high curvature to the boundary cannot be neglected. Hence, the kink-induced-directionality is lost.\\
  
  The crossover from linear to circular motion (discussed below) occurs between obtuse $\theta_{kink}=\frac{5\pi}{6}$ and acute angles $\theta_{kink}=\frac{\pi}{2}$. While for most trimer systems the transition shifts directly from linear forward to linear backward motion, we observe an additional regime of predominantly linear backward-moving vesicles for dimer systems and trimer systems with low activity. Here we distinguish between two mechanisms. For $0.1<\rho_{N_a}<0.15$, the geometry of the kink, while still inducing clustering, prevents the stability of large parallel stackings. Hence, in some of our simulations this geometric frustration destabilizes rod alignment either completely (see the symbol $\times$ in Fig.\ref{fig:results2}) yet does not break up the cluster completely, or partially shifts it to the opposite side of the vesicle. Here the boundary features a second local curvature maximum and the parallel rod packing is more stable. This causes the forward motion of vesicles to become less reliable with decreasing kink angle and eventually to flip to backward linear motion for kink angles close to $\theta_\text{kink}=\frac{7\pi}{12}$. We observed a similar phenomenon in the trimer-filled vesicles with increasing particle density, but the effect is not as strong as in the dimer systems and only occurs at lower $Pe=25$ and over a narrower region of $\frac{\pi}{2} \leq \theta_\text{kink} \leq \frac{2\pi}{3}$.\\
  
  The second mechanism that reverses the direction of motion of the dimer-filled vesicle occurs for $\rho_{N_a}>0.15$. Here the dimers form two clusters, one located at the kink and the other located at the opposite side of the vesicle. For large obtuse $\theta_\text{kink}$ the cluster at the kink is larger, effectively pushing the vesicle forward with high reliability. For smaller $\theta_\text{kink}$ and less stable alignment of rods at the kink, the cluster of rods at the opposite side of the vesicle becomes larger, eventually dominating and reversing the direction of motion of the vesicle superstructure. The two counteracting clusters also explain the decrease in the velocity of the dimer-filled vesicles (see Fig.~\ref{fig:results2}a) best observed in the velocity distribution in Fig.~\ref{fig:results3}a, where both fast vesicles with one dominant cluster (peak at high velocity) and slow vesicles with two counteracting clusters (peak close to $v=0$) are present. In contrast, trimer-filled vesicles, where only one large cluster forms, move faster with more contributing active agents.\\
  
  \subsection{Circular motion}
  
  In addition to linear motion, vesicle superstructures also exhibit circular motion where they rotate with a mean angular velocity $\langle \dot{\omega}\rangle$ around the superstructure's center of mass while moving linearly either forward or backward. While circular motion can also, in principle, occur from an unbalanced distribution of parallel stacked particles around the kink, that mechanism causes only a slight bend in the vesicle's trajectory. The radii of most circular trajectories with \textit{parallel} stacked particles are larger than 500$\sigma$, which are perceived as linear motion over our simulation time. In most cases in which vesicles move with higher angular velocity, we observe a different stacking behavior of the active particles. Rather than \textit{parallel}, the rods exhibit \textit{angled} stacking near the kinked tip. In circularly moving vesicles with acutely angled kinks ($\theta_\text{kink}>\frac{\pi}{2}$) the perpendicular velocity component stems from an asymmetrically applied pressure at the kink (see sketch in Fig. \ref{fig:methods}b). As the likelihood of the rods unevenly occupying the space around the kink increases with $\rho_{N_a}$ and decreases with $\theta_\text{kink}$, the angular velocity increases in this region of phase space.\\
  
  For vesicles with $\theta_\text{kink}<\frac{\pi}{2}$ the direction of the rod alignment changes. Here, the angle is sufficiently narrow that the first trapped rod is pushed to the boundary by all other incoming particles. As a consequence, the first rod forms a new, effective "kink" with the vesicle boundary where another rod can settle. This mechanism cascades to produce an array of rods that are all aligned with one side of the vesicle wall (see Fig.~\ref{fig:methods}). Due to the resulting offset in the alignment angle $\alpha\approx\frac{\theta_\text{kink}}{2}$ from the symmetry axis (see third column in Fig.~\ref{fig:results3}) we can decouple the applied forces at the kink into a parallel and a perpendicular force component, resulting in a net circular motion of the vesicle. The long stacking of particles at one side of the kink is also apparent in the particle distribution in Fig.~\ref{fig:results2}, where the peaks are more stretched along the vesicle boundary for $\theta_\text{kink} < \frac{\pi}{2}$ than for $\theta_\text{kink}>\frac{\pi}{2}$. Compared to linearly moving vesicle superstructures, however, both the velocity and angular velocity distribution are wider and increase with $\rho_{N_a}$. The kink-forward and kink-backward circular regime depends on the existence of a second cluster, which counteracts the forward pushing component but not the rotational component of the stacking. Hence, only dimer-filled vesicles can achieve a kink-backward circular motion dominant regime as shown in Fig.~\ref{fig:results1}.  \\
  
\begin{figure*}
      \centering
      \begin{subfigure}[b]{0.8\linewidth}
      \centering
          \includegraphics[width=\linewidth]{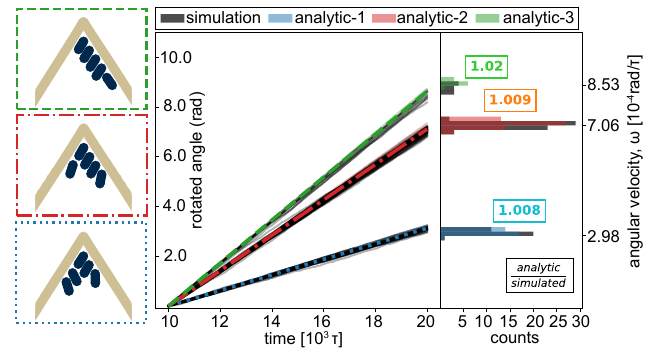}
          \caption{$\alpha=2$, $Pe=75$}\label{fig:results4a}
      \end{subfigure}
      \begin{subfigure}[b]{0.8\linewidth}
          \centering
          \includegraphics[width=\linewidth]{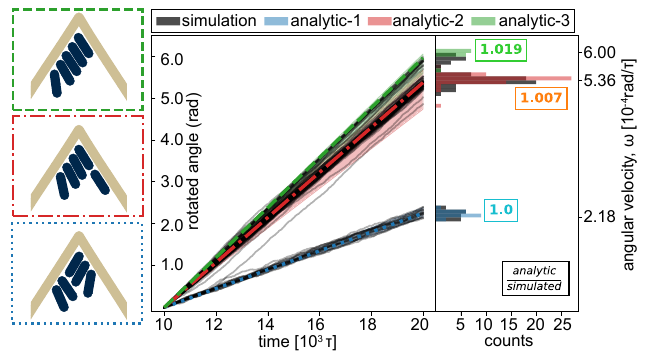}
          \caption{$\alpha=3$, $Pe=50$}\label{fig:results4b}
      \end{subfigure}
      \caption{\label{fig:results4} \textbf{Analytic calculation of angular velocities from different stacking arrangements.} From the simulation of vesicles with $\theta_{kink} = \pi/3$ and $\rho_{N_a} = 0.05$, the rotated angle of the vesicle from its initial orientation is plotted against time. Each black line is from a single simulation trajectory. On the right, histogram of angular velocities from each instance is plotted in black. Sketches on the left are stacking arrangements of active particles observed at the kinked tip of the vesicle. The analytical angular velocities are calculated from each simulation replica, grouped by each stacking sequence, and plotted in the histogram with matching color to the border of the insets. Finally, the rotation angle plotted against time of three different stackings of rods is interpolated from the analytical angular velocity and drawn with matching color and line style to the insets. The values squared in the histogram are the ratio of the analytical angular velocity to the simulated angular velocity, averaged over replicas.}
  \end{figure*}

  To relate the stacking of the particles to the angular momentum of the vesicle superstructure in more detail, we predict its angular velocity by analytically calculating the rotation of different ideal stacking sequences in the vesicle. For convenience we focus only on one specific vesicle system in the following ($\theta_\text{kink}=\pi/3$, $\rho_{N_a}=0.05$), although the same approach also applies to all other vesicle shapes and rod densities. In this system with five active rods we can build three different stackings that cause a rotation (see Fig.\ref{fig:results4}). Each stacking sequence represents the three different ways of aligning the active rods with both sides of the kink, considering symmetric equivalence. For each particle $i$ included in the vesicle transfers its linear momentum  to the vesicle such that, $ \sum_{i=1}^n |\Vec{v}_{i,\bot}|\cdot |l_i| = I_{vesicle} \cdot \omega $, where $l_i$ is the position vector of particle $i$ in relation of the center of mass of the vesicle to the particle $i$, $\Vec{v}_{i,\bot}$ is the active velocity of particle $i$ perpendicular to $l_i$ and $I_{vesicle}$ is the moment of inertia of the vesicle.\\
  
  For comparison we obtain from the simulation the rotation angle of the vesicle from its initial orientation at each time step (black bars in Fig.~\ref{fig:results4}) and the corresponding angular velocities (color bars in Fig.~\ref{fig:results4}). We observe that the angular velocities of each simulated vesicle fall into one of three groups. Each group corresponds to one of the ideal stacking sequences, indicated by the agreement in angular velocity for both dimer and trimer systems. As we introduce more rod particles into the vesicle, additional types of stacking sequences can be constructed, resulting in the splitting of the angular velocities into even more groups and a wider distribution of the angular velocity as shown in Fig.\ref{fig:results3}.
  
  \subsection{switchable motion}
\begin{figure*}
  \centering
  \includegraphics[width=0.95\linewidth]{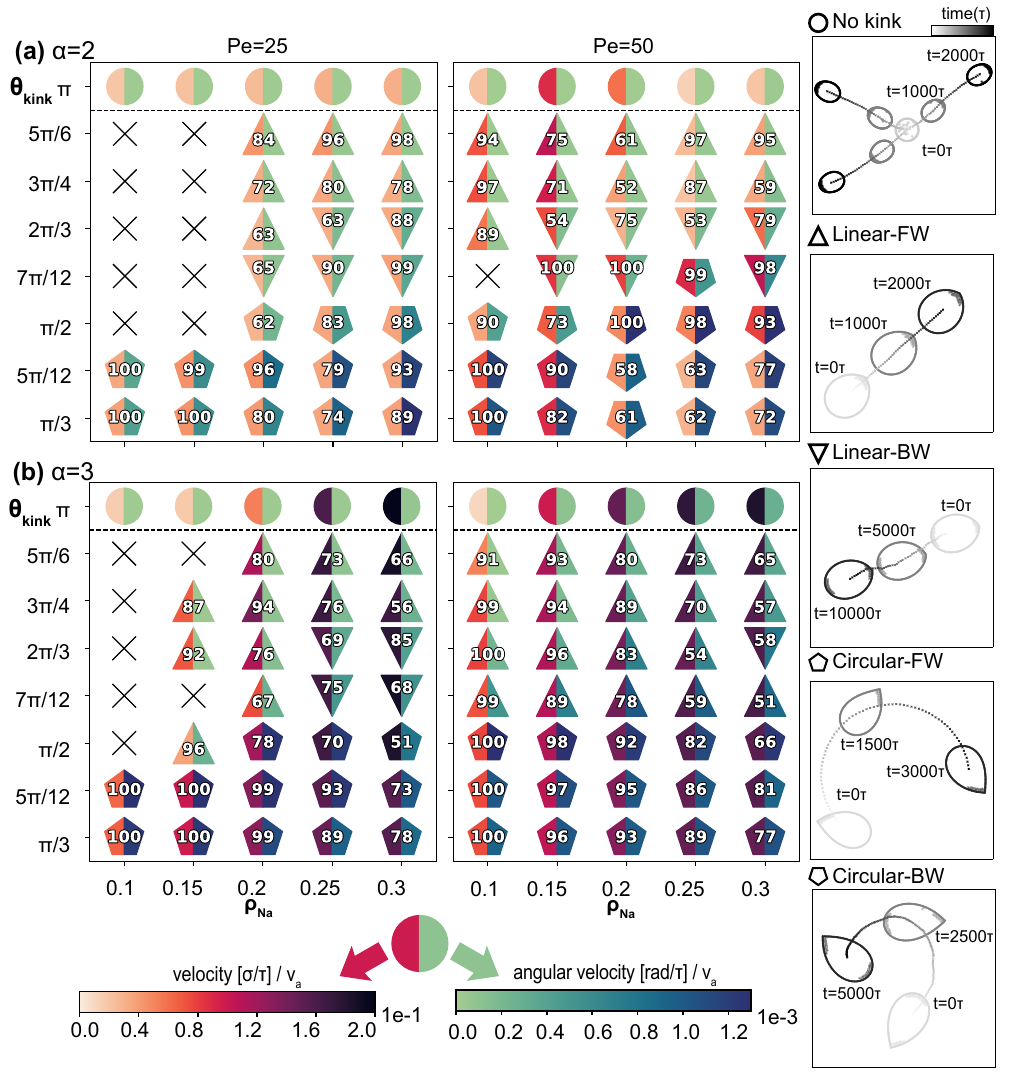}
  \caption{\label{fig:results1} \textbf{Chart of directional motion of the vesicle with regards to $\theta_{kink}$ and $\rho_{N_a}$} The right column is the legend for the left four charts, and describes the statistically dominant motions of the vesicle superstructures with and without a kink. The markers indicate unbiased motion ($\bigcirc$) and the four newly observed types of motion: linear forward (FW) motion ($\triangle$), linear backward (BW) motion ($\bigtriangledown$), circular forward motion (\protect\pentagonup), and circular backward motion (\protect\pentagondown)). The left four steady-state charts are diagrams that indicate regions of the different behaviors. The symbol $\times$ indicates that the majority of vesicles with active rods have not clustered within the simulation time. Linear and circular motion are sorted based on the stacking arrangement of the enclosed active particles across the range of $\theta_{kink}$ and $\rho_{N_a}$ studied. (Note that $\theta_{kink} = \pi$ corresponds to vesicles without a kink). The color of the left and right halves of each marker indicates the velocity of the vesicle normalized by the active velocity $v_a$ (left color bar), and the angular velocity of the vesicle (right color bar), respectively. The number on each marker denotes the percentage of vesicles that move according to the marker. The remaining percentage of vesicles moves opposite to that indicated. Vesicles with $\theta_{kink}=\pi$ have numberless markers because there is no kink axis. Each data point (marker) is averaged over 200 independent replicas. \textbf{(a) $\alpha=2$, (b) $\alpha=3$}}
  \end{figure*}

  All of the observed dynamical behaviors of the active vesicle superstructures emanate from the spontaneous alignment of rods into clusters at the kink in the vesicle boundary.  If this propelling mechanism is to be used to navigate and control the superstructure, we must be able to switch among the various behaviors dynamically and reversibly. To test this, we perform three sets of simulations that represent a horizontal (particle density), vertical (kink angle) and loop (both density and kink angle in sequence) dynamical change in regard to the steady-state diagram in Fig.~\ref{fig:results1}.\\ 
  
  To alter $\rho_{N_a}$ during our simulation we introduce two types of active rods with similar shape that can be activated independently. By only turning on one type of trimer confined within a vesicle with $\theta_\text{kink} = \frac{5\pi}{6}$ we first observe the active particles forming the parallel stacking that induces kink-forward linear motion while the still passive trimers are dragged as cargo at the opposite end of the vesicle. Once we activate the second rod type, the newly self-driven trimers join the existing rod packing at the kink, adding to the propulsion and accelerating the superstructure. The vesicle returns to its regular speed as soon as one active rod type is turned off, breaks away from the cluster and accumulates at the back of the vesicle as cargo again (see video 1). By repeating this activation sequence with dimer particles we observe that the vesicles slow down, stop their linear forward motion, or even exhibit backward motion before they obtain their original speed again (see video 2). This indicates that the speed of the vesicle is controllable and reversible through the creation and destruction of counteractive clusters of rods.\\
  
  Vesicles that continuously change their kink angle from $\theta_\text{kink} = \frac{\pi}{2}$ to $\theta_\text{kink} = \frac{5\pi}{6}$ and back, dynamically switch between circular and linear motion (see videos 3 and 4). As predicted by the steady-sate diagram in Fig.~\ref{fig:results1}, the parallel stacking of active rods in the linear forward motion regime becomes geometrically unstable or pushed to one side during the kick angle narrowing process. Hence the dimers and trimer rearrange from aligning parallel to the vesicle symmetry axis to aligning at an angle, causing subsequent circular motion of the vesicle. By widening the angle again the particles can slide from one side of the wall into the center and recover the original parallel stacking and associated linear motion.\\
  
  Although the steerability of the vesicle is most reliable in the low density regime $\rho_{N_a} < 0.25$, we can combine both the adjustable angle mechanism and the multiple active particle type mechanism to trigger a variety of preprogrammed superstructure locomotion. Video 5 shows a vesicle filled with trimers that first moves linearly (one active/one passive type + obtuse kink angle), increases its speed (two active types + obtuse kink angle), takes a turn (two active types + acute kink angle), reduces its velocity to the original speed (one active/one passive type + acute kink angle) and finally exhibits a slow linear trajectory again. A similar loop in the phase space of vesicles filled with dimers results in a sequence of forward circling (one active/one passive type + acute kink angle), backward circling (two active types + acute kink angle), backward linear motion (two active types  + obtuse kink angle), forward linear motion (one active/one passive type + obtuse kink angle) and finally returning to forward circling. The trend in directionality and spatial distribution of active particles that favors the kinked area of the boundary weakens as the perimeter of the vesicle increases.
  
\section{Discussion}
In this paper, we used Brownian molecular dynamics simulations to study the rich dynamical behavior of rigid kinked vesicles that contain self-propelling rod-shaped particles. We identified that kinks in the vesicle membrane bias the emergent clustering and alignment of the active agents. Based on the angle of the kink, density of the active particles, and the active particle length, the rods form different stacking sequences at or opposite to the kink and collectively induce multiple types of directed motion by the vesicle superstructure. Specifically, we identified a kink-forward linear motion, a kink-backward linear motion, a kink-forward circular motion and a kink-backward circular motion. 
By analytically calculating the rotation of different rod stacking sequences in the vesicle, we showed how rod packing at and around the kink correlates with the resulting superstructure motion.  Finally, we demonstrated that our model allows for dynamic interchangeability between the different types of motion and adjustable superstructure speed by changing the vesicle geometry or active particle properties.\\

Our findings suggest that both microscopic and macroscopic vesicular superstructures could be controlled by just two steering parameters, the geometry of the kinked vesicle and the number of active rods, bringing us closer to realizing autonomous robotic entities. Motile bacteria or active colloids encapsulated inside of a giant lipid vesicle are known to impart motion to the entire vesicle \cite{le_nagard_encapsulated_2022}. Including anisotropic elements with controllable position inside the vesicle membrane \cite{xin_switchable_2021-1}, or engulfing mixtures of active particles that respond to different external stimuli \cite{niese_apparent_2020, salinas_lorentz_2021, ziepke_multi-scale_2022}, could open the door for artificial cell systems that display directional movement for targeted delivery or retrieval of micron-scale or smaller objects. Similarly, in macroscopic machines, such as the superstructure reported in \cite{boudet_collections_2021}, implementing adjustable bending stiffness to the confining wall, either mechanically or via actuators \cite{guo_adjustable_2021}, could be the key to guide the location and collective alignment of the individual active machines and to realize smarter, more adaptable swarm robots.

\section{Acknowledgements}
This work was supported as part of the Center for Bio-Inspired Energy Science, an Energy Frontier Research Center funded by the U.S. Department of Energy, Office of Science, Basic Energy Sciences under Award \# DE-SC0000989. This work used the Extreme Science and Engineering Discovery Environment (XSEDE) \cite{towns_xsede_2014}, which is supported by National Science Foundation grant number ACI-1548562 (XSEDE award DMR 140129); computational resources and services were also provided by Advanced Research Computing (ARC), a division of Information and Technology Services (ITS) at the University of Michigan, Ann Arbor.

\clearpage
\bibliography{references}

\input{sections/6appendix.tex}

\end{document}

%% file: sections/6appendix.tex
\section{Supplementary} 

\subsection*{Two dimensional probability density map of finding a rod in the vesicle}
We use kernel density estimation (KDE) with a Gaussian kernel to estimate the probability distribution of rods' relative location in the vesicle in the steady state. The probability to find a rod is measured per rectangular bin located at $(w_i, h_j),$ where $i,j \in [1,100]$. The bins are equally spaced in the rectangular region that spans between the lower and upper limit of the width of the vesicle along the $w$ direction and of the height of the vesicle along the $h$ direction. 

\subsection*{Shape of the vesicle with respect to bending stiffness}
The vesicle shape in steady state varies with vesicle stiffness. The vesicle stiffness is set by the harmonic angle "spring constant", $\kappa_{a}$, applied to every neighboring triplet of disks that composes the vesicle. The shape elongation is determined by eigenvalues and eigenvectors of the vesicle boundary gyration tensor. In two dimensions, the x-y component of the gyration tensor is

$s_{xy} = \frac{1}{2N^2} \sum_{i=1}^N \sum_{j=1}^N (x_i-x_j)(y_i-y_j)$

The square roots of the eigenvalues ($\sqrt{L1}, \sqrt{L2}$) are the characteristic principal-axis lengths (radii) of the ellipsoid that describesthe shape of the vesicle. The metric for the shape descriptor is defined as a ratio of the shortest gyration moment to the longest gyration moment: $Ratio(L1,L2)=\frac{\sqrt{L1}}{\sqrt{L2}}$. The ratio is plotted in Fig.~\ref{fig:suppl-shape} for each rod aspect ratio ($\alpha = 2,3$) with various $\kappa_{a}=50,125,250,500,1000k_bT$, $\theta_{kink}=\pi/2,5\pi/6$, and number densities $\rho_{N_a}=0.1,0.15,0.2$. The value of the shape descriptor decreases with decreasing vesicle stiffness, which means that the vesicle is more elongated with more flexible boundaries. The reference points marked with black squares are from the vesicles containing no propelling rods. The shape descriptor deviates more from the reference point as vesicle stiffness decreases. The higher number density $\rho_{N_a}$ leads to more elongation.

\renewcommand{\thefigure}{S\arabic{figure}}
\setcounter{figure}{0}

\begin{figure}[!htbp]
\centering
\includegraphics[width=0.95\linewidth]{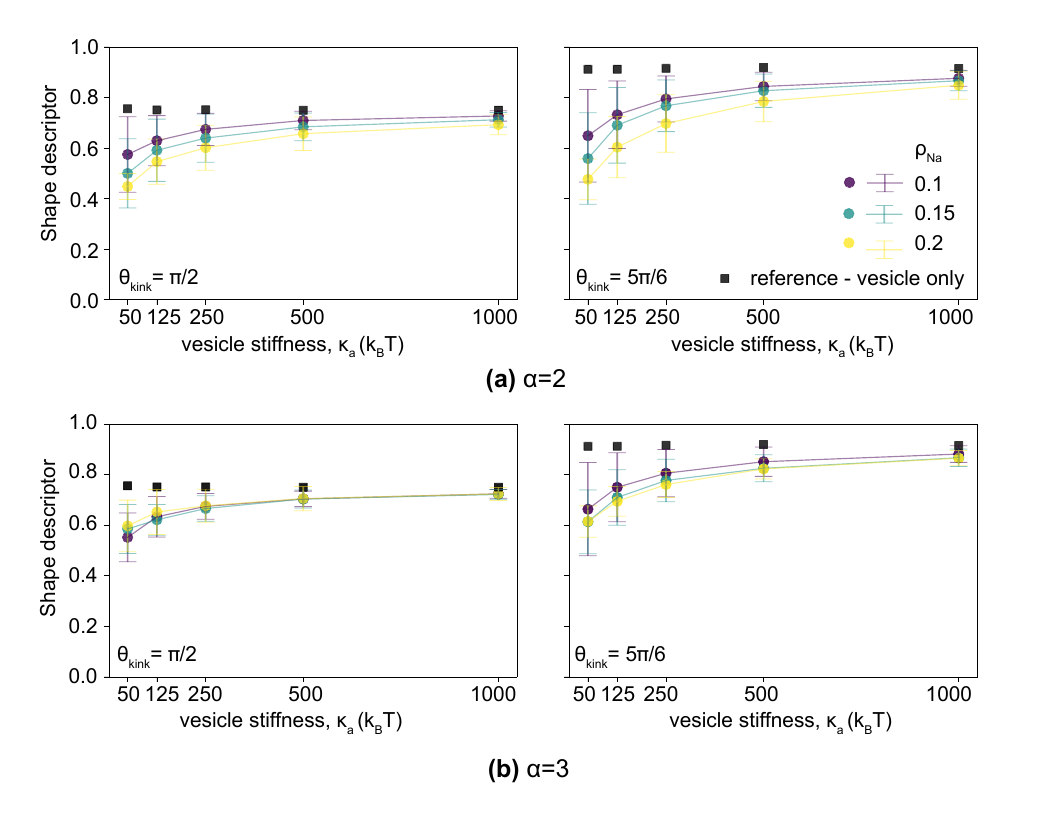}
\caption{\textbf{Shape elongation of the vesicle.}  Shape descriptor measure of the extent of elongation for different values of vesicle bending stiffness $\kappa_{a}$ and different aspect ratios $\alpha$ of the rod. Color refers to the number densities $\rho_{N_a}=0.1,0.15,0.2$. The marker is the averaged descriptor value with error bar of matching color. The reference system for each value of bending stiffness is from a vesicle with no active particles.}
\label{fig:suppl-shape}
\end{figure}

\begin{figure*}[!htbp]
\centering
\includegraphics[width=0.9\linewidth]{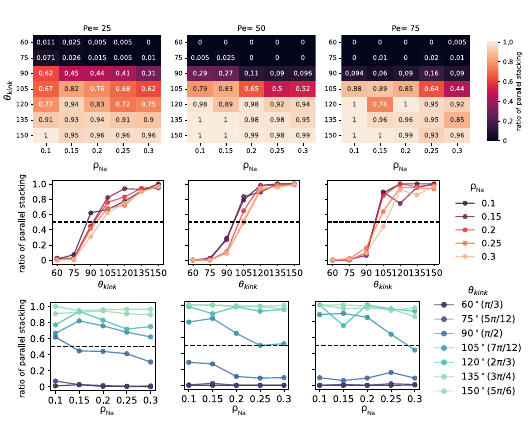}
\caption{\textbf{Ratio of parallel vs. angled stacking plotted over vesicles with $\alpha=2$ for each data point with corresponding Pe, $\theta_{kink}$ and $\rho_{N_a}$.} For each moving vesicle the stacking was measured either $\textit{parallel}$ or $\textit{angled}$. Angled stacking corresponds to cases when the closest active rod particle to the kink aligns with the vesicle boundary; otherwise the rods are considered to be stacked in parallel.}
\label{fig:suppl-stackingcode-dimer}
\end{figure*}

\begin{figure*}[!htbp]
\centering
\includegraphics[width=0.9\linewidth]{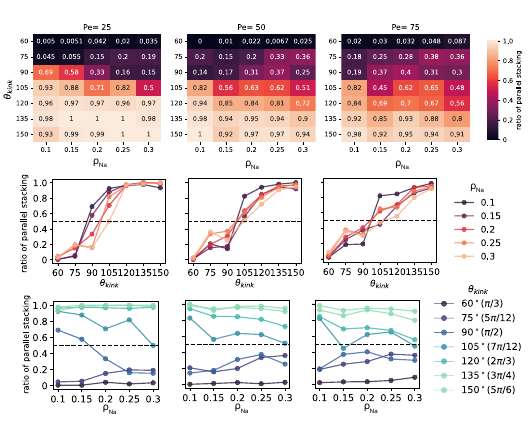}
\caption{\textbf{Ratio of parallel vs. angled stacking plotted for vesicles with $\alpha=3$ for each data point with corresponding Pe, $\theta_{kink}$ and $\rho_{N_a}$.} The measurement method is identical to that in Fig.~S2. }
\label{fig:suppl-stackingcode-trimer}
\end{figure*}

\begin{figure*}[!htbp]
\centering
\includegraphics[width=0.9\linewidth]{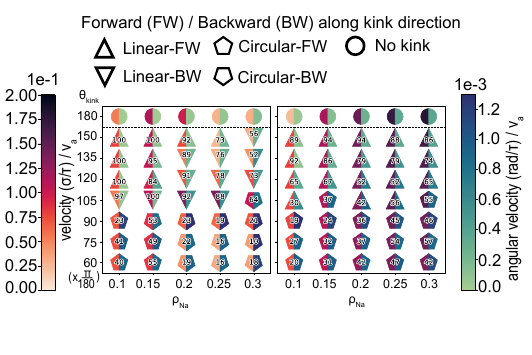}
\caption{\textbf{Chart of directional motion of the vesicle with regards to Pe=75, $\theta_{kink}$ and $\rho_{N_a}$.} The color scheme is the same as in Fig.~\ref{fig:results1}}
\label{fig:suppl-phase-diagram-75}
\end{figure*}

\begin{figure*}[!htbp]
\centering
\includegraphics[width=0.9\linewidth]{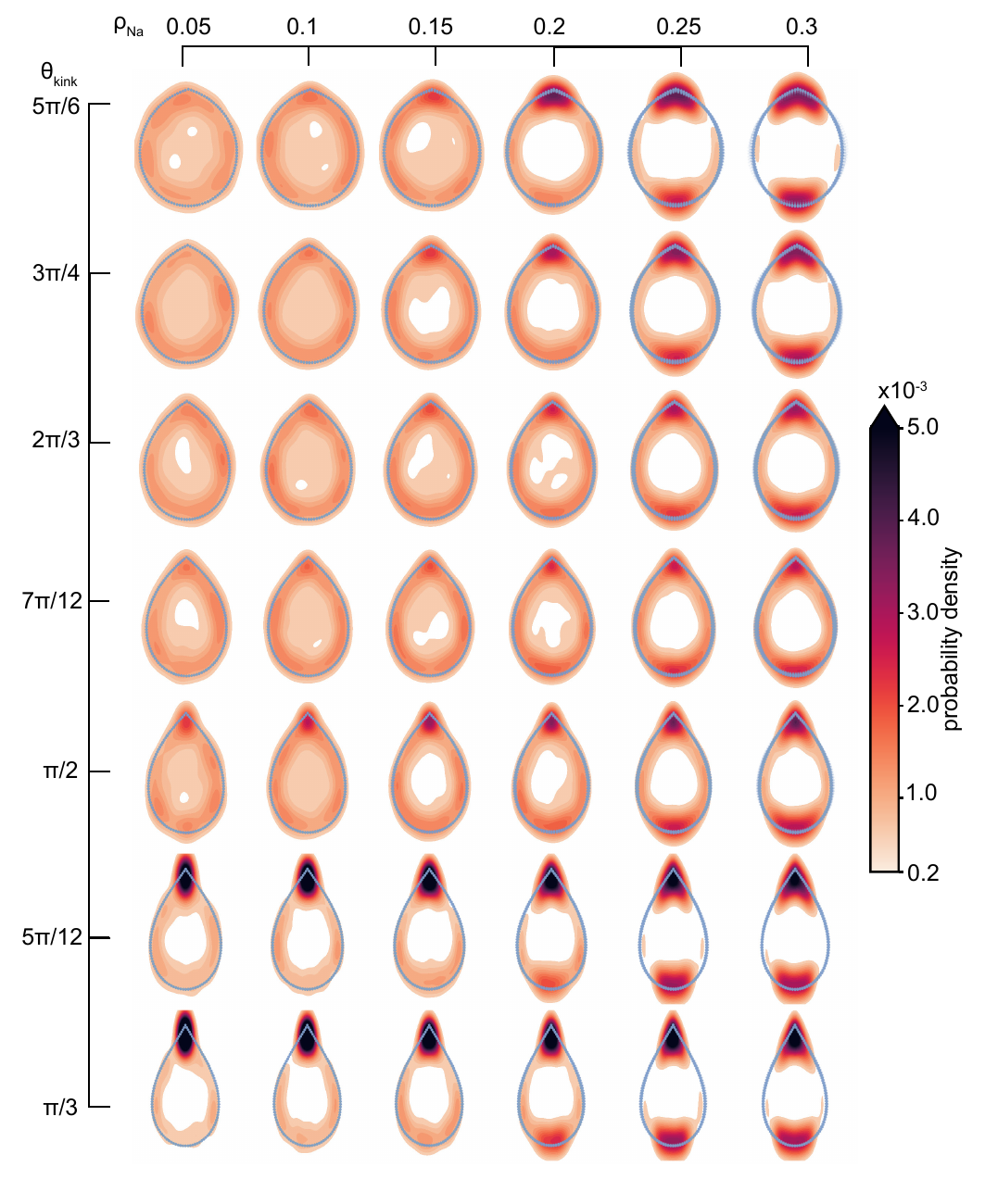}
\caption{\textbf{Probability density plot of enclosed particle $\alpha=2$, $Pe=25$, $\kappa_{a}=1000k_BT$.} The color scheme is the same as in Fig.~\ref{fig:results2}}
\label{fig:suppl-kde1}
\end{figure*}

\begin{figure*}[!htbp]
\centering
\includegraphics[width=0.9\linewidth]{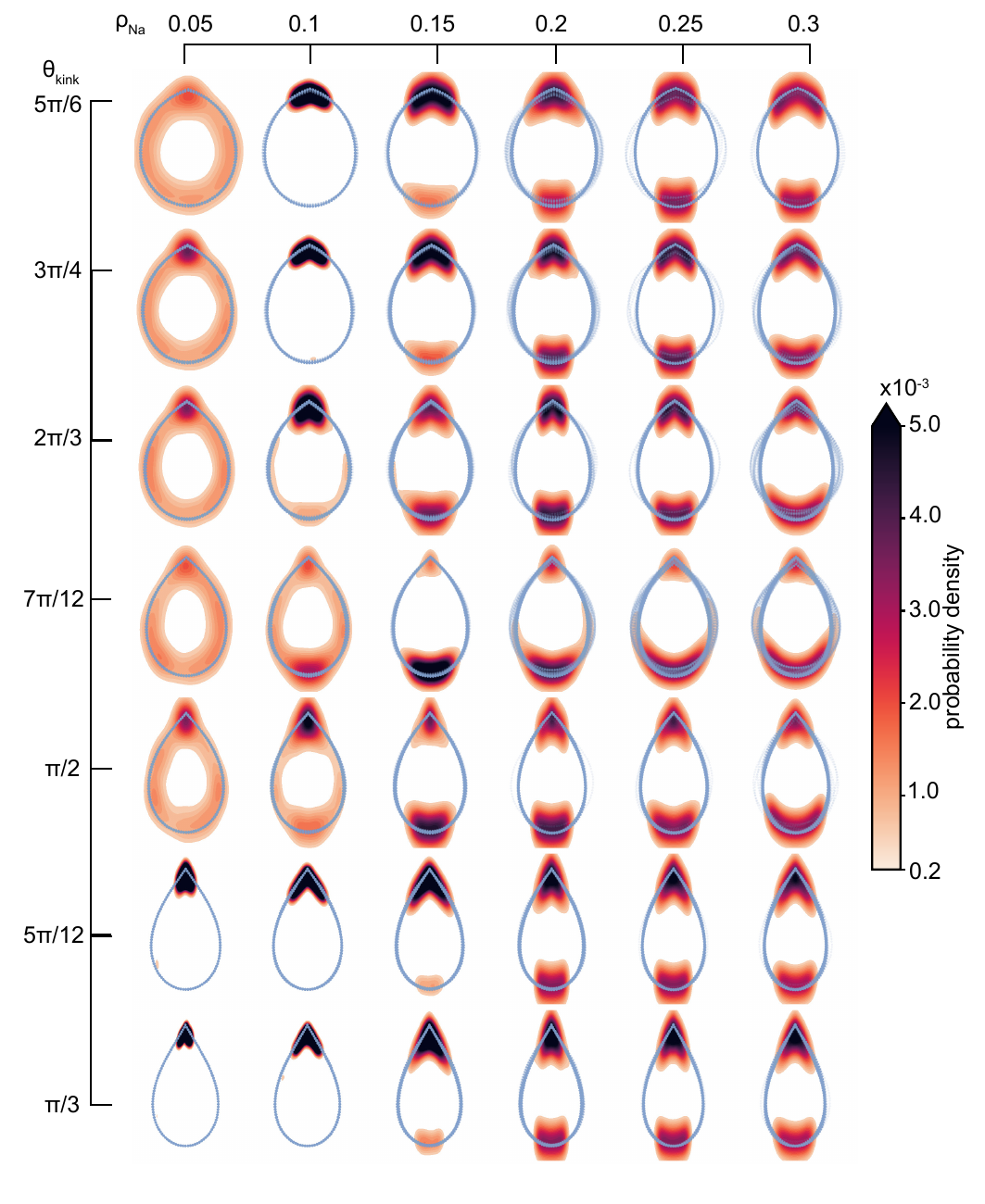}
\caption{\textbf{Probability density plot of enclosed particle $\alpha=2$, $Pe=50$, $\kappa_{a}=1000k_BT$.} The color scheme is the same as in Fig.~\ref{fig:results2}}
\label{fig:suppl-kde2}
\end{figure*}

\begin{figure*}[!htbp]
\centering
\includegraphics[width=0.9\linewidth]{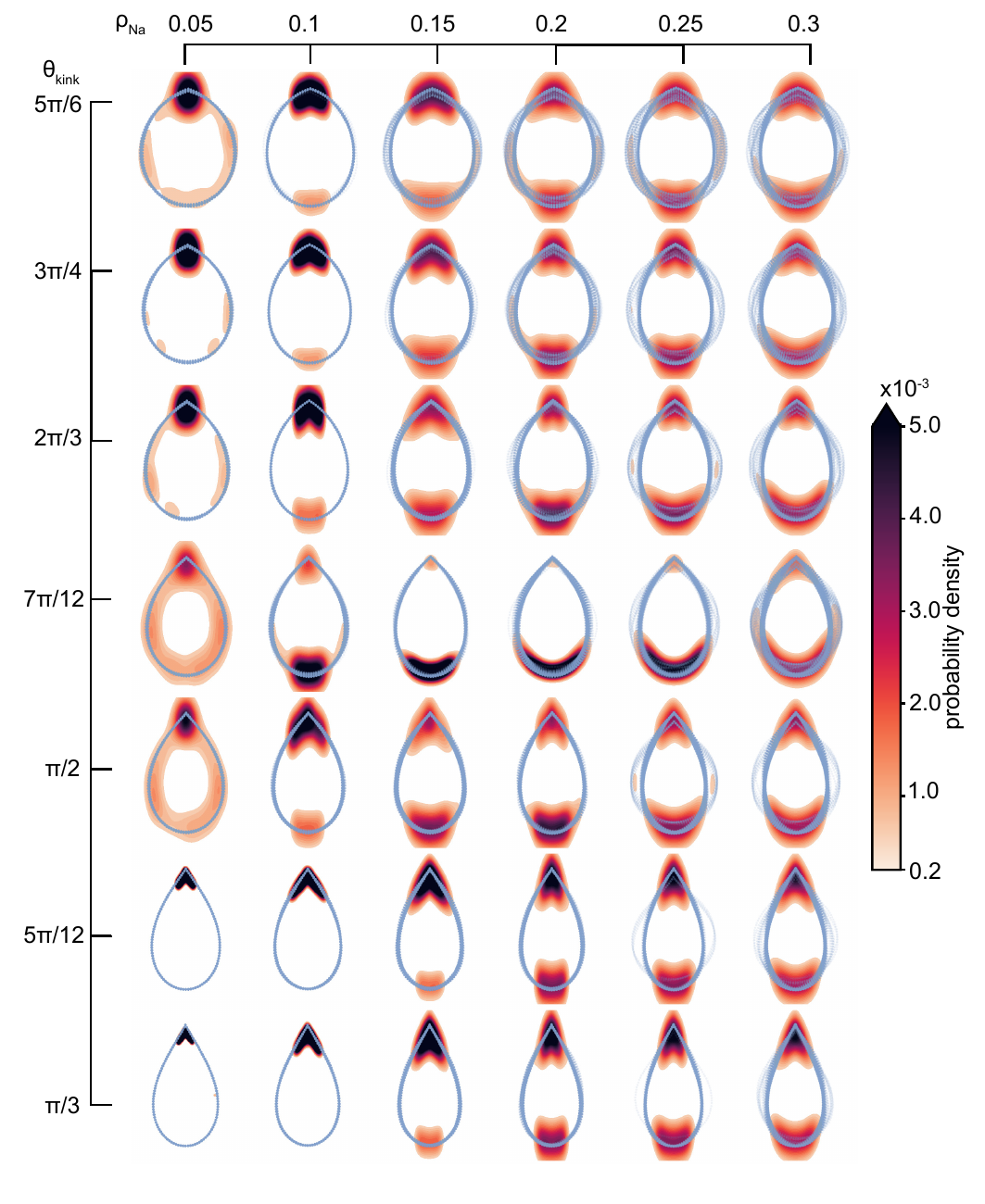}
\caption{\textbf{Probability density plot of enclosed particle $\alpha=2$, $Pe=75$, $\kappa_{a}=1000k_BT$.} The color scheme is the same as in Fig.~\ref{fig:results2}}
\label{fig:suppl-kde3}
\end{figure*}

\begin{figure*}[!htbp]
\centering
\includegraphics[width=0.9\linewidth]{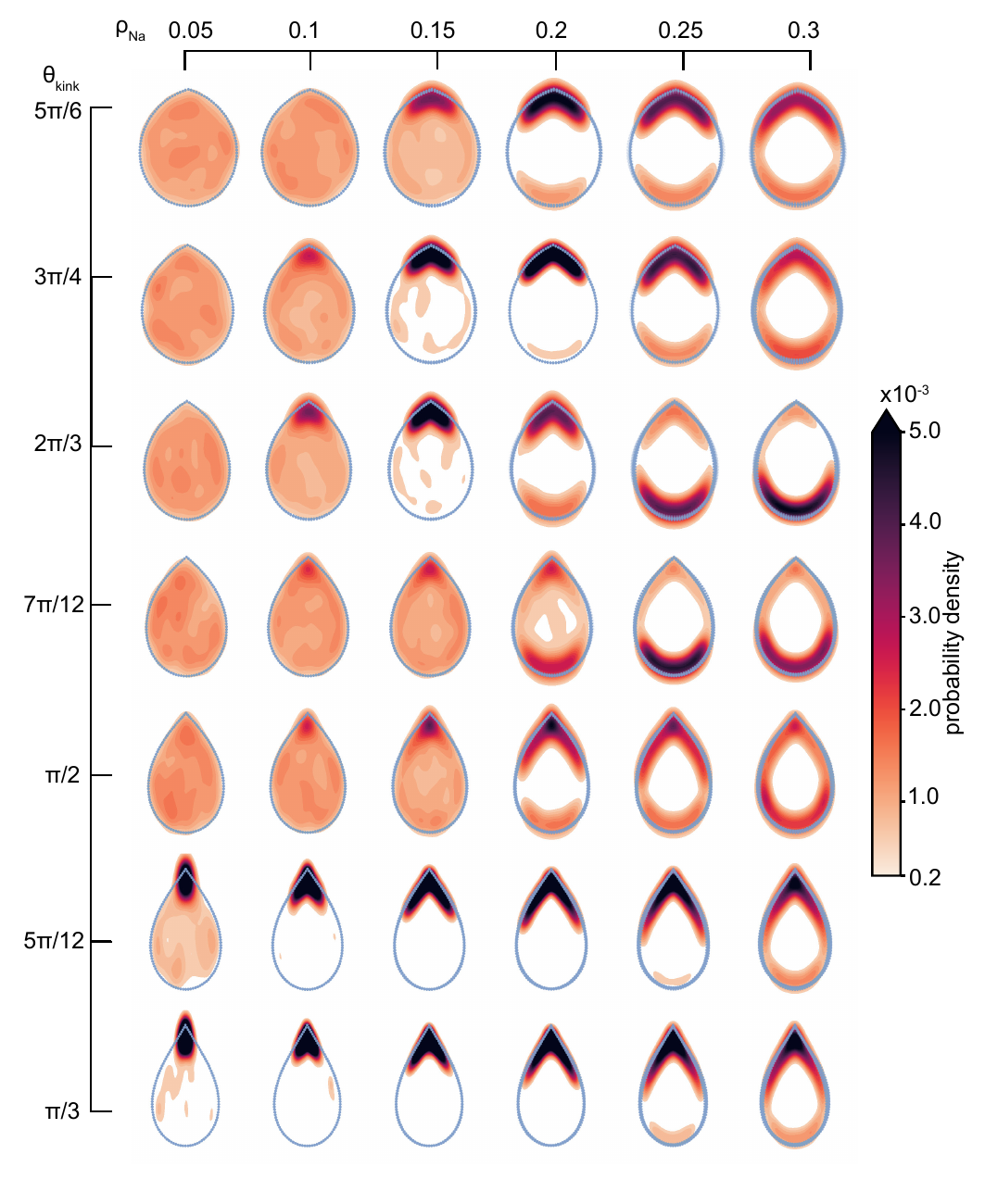}
\caption{\textbf{Probability density plot of enclosed particle $\alpha=3$, $Pe=25$, $\kappa_{a}=1000k_BT$.} The color scheme is the same as in Fig.~\ref{fig:results2}}
\label{fig:suppl-kde4}
\end{figure*}

\begin{figure*}[!htbp]
\centering
\includegraphics[width=0.9\linewidth]{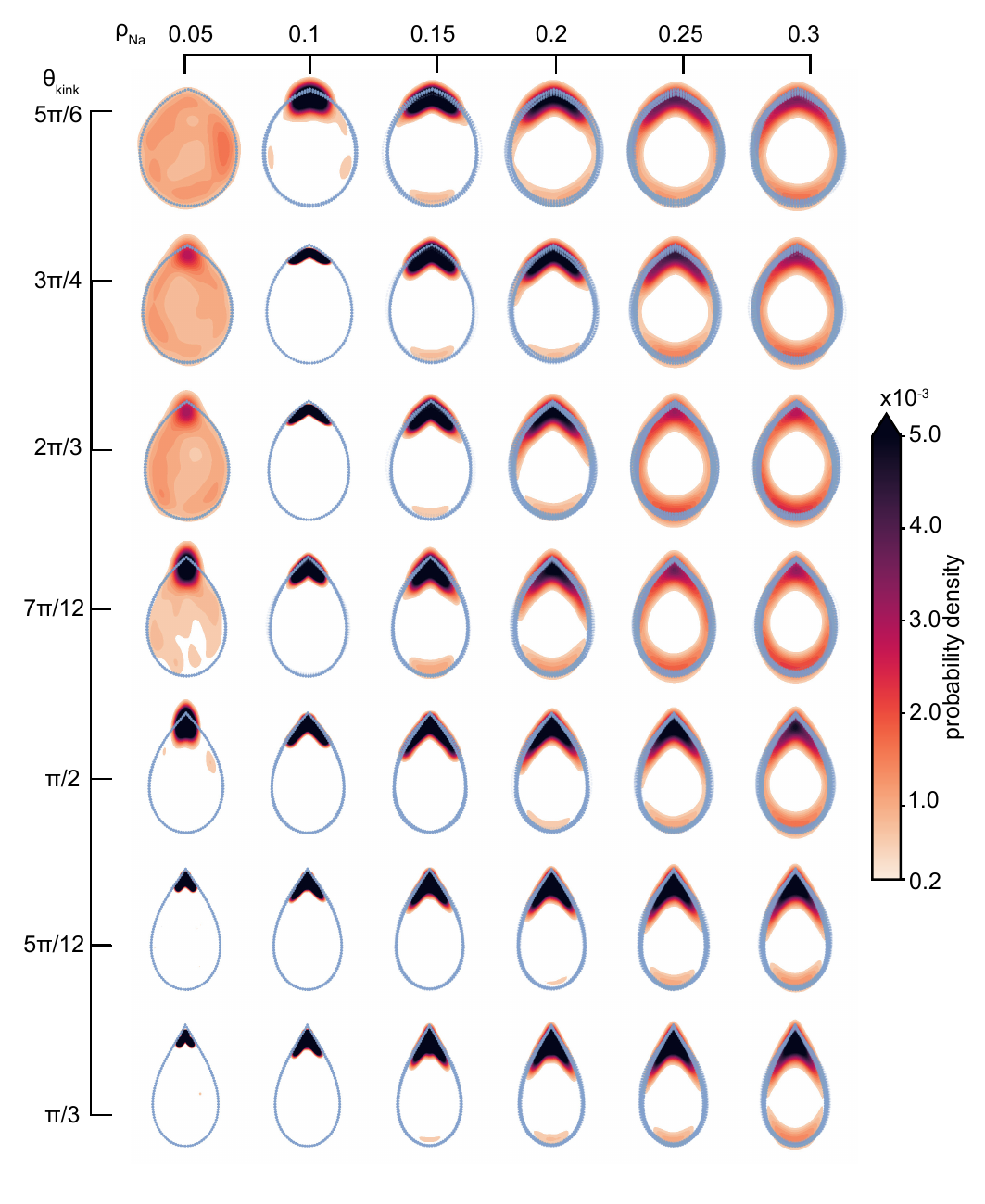}
\caption{\textbf{Probability density plot of enclosed particle $\alpha=3$, $Pe=50$, $\kappa_{a}=1000k_BT$.} The color scheme is the same as in Fig.~\ref{fig:results2}}
\label{fig:suppl-kde5}
\end{figure*}

\begin{figure*}[!htbp]
\centering
\includegraphics[width=0.9\linewidth]{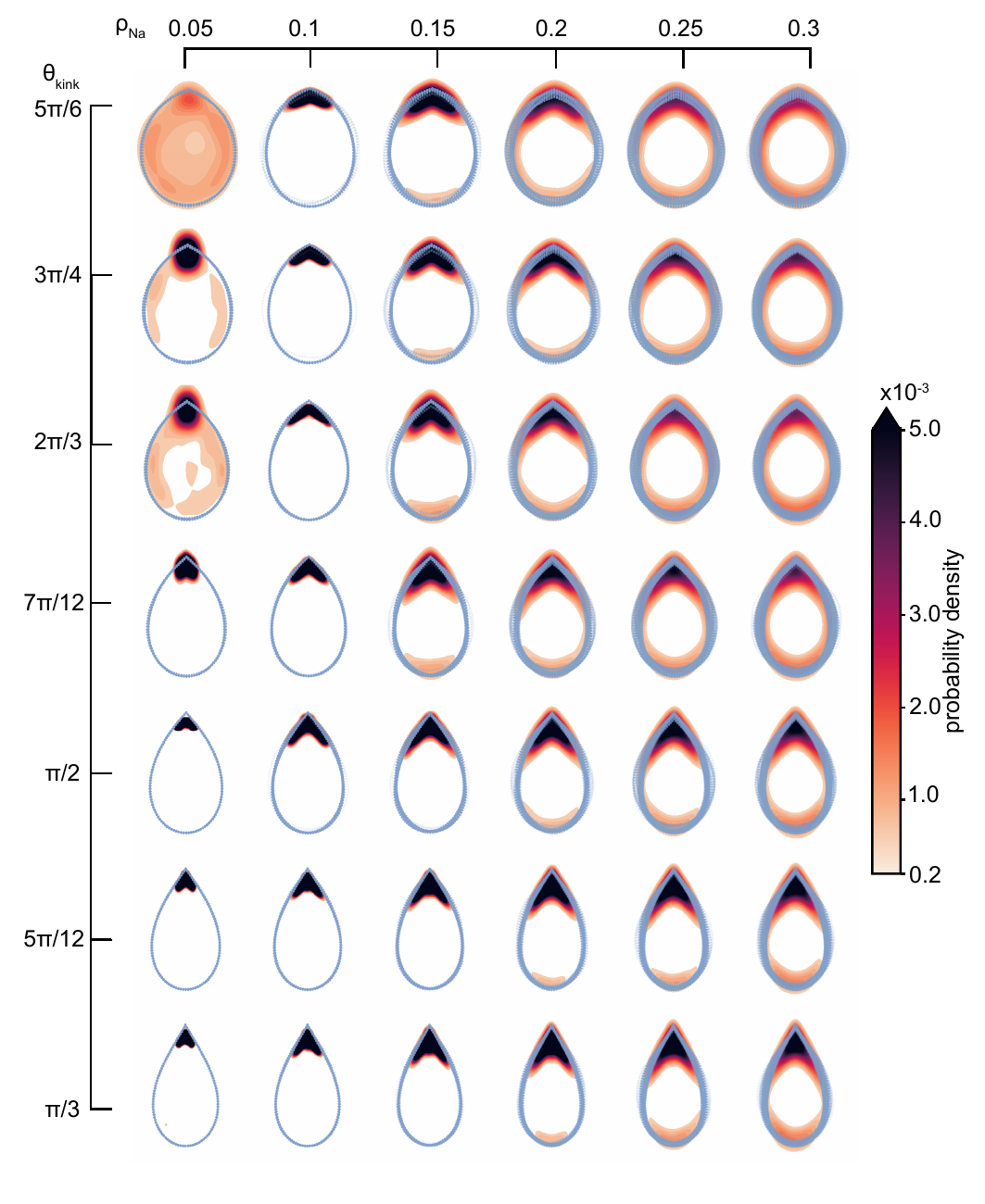}
\caption{\textbf{Probability density plot of enclosed particle $\alpha=3$, $Pe=75$, $\kappa_{a}=1000k_BT$.} The color scheme is the same as in Fig.~\ref{fig:results2}}
\label{fig:suppl-kde6}
\end{figure*}

\begin{figure*}[!htbp]
\centering
\includegraphics[width=0.9\linewidth]{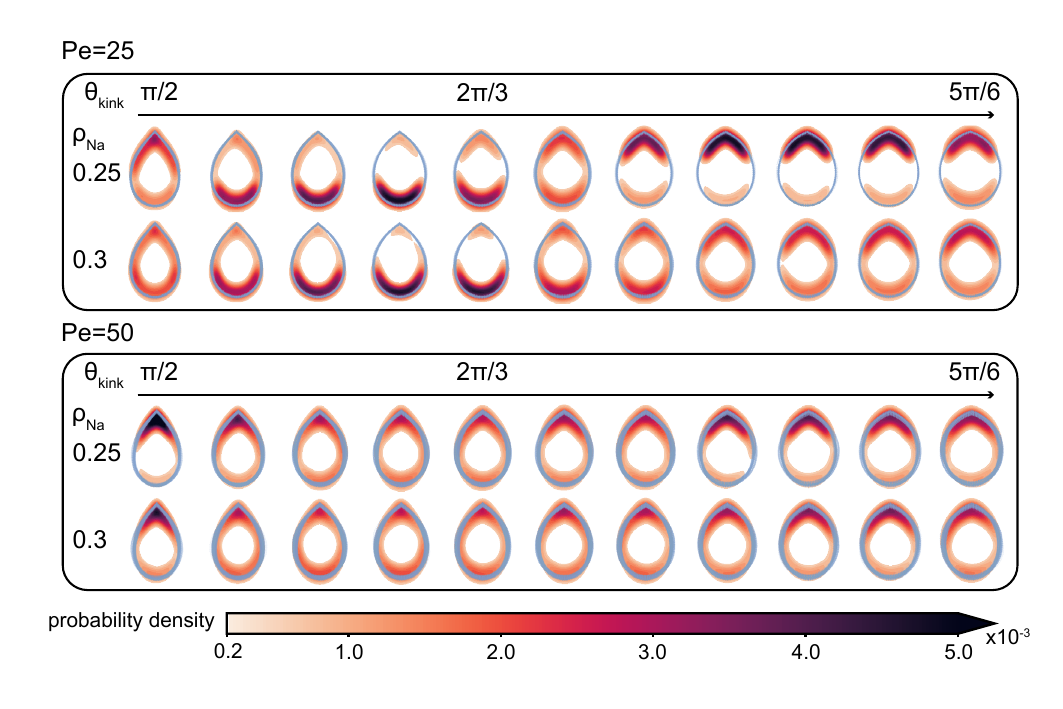}
\caption{\textbf{Flipped trend in probability density plot of enclosed particle with various $\theta_{kink}$ $\alpha=3$, $Pe=25, 50$, $\kappa_{a}=1000k_BT$.} The color scheme is the same as in Fig.~\ref{fig:results2}}
\label{fig:suppl-flipped-trend}
\end{figure*}